# Probing the Origin of Extreme Magnetoresistance in Pr/Sm Mono-Antimonides/Bismuthides

Zhongzheng Wu[1#], Fan Wu[1#], Peng Li[1], Chunyu Guo[1], Yi Liu[2], Zhe Sun[2], Cheng-Maw Cheng[3], Tai-Chang Chiang[4], Chao Cao[5], Huiqiu Yuan[1,6*], and Yang Liu[1,6*]

[1]Center for Correlated Matter and Department of Physics, Zhejiang University, Hangzhou, P. R. China

[2]National Synchrotron Radiation Laboratory, University of Science and Technology of China, Hefei, P. R. China

[3]National Synchrotron Radiation Research Center, Hsinchu, Taiwan, Republic of China

[4]Department of Physics and Frederick Seitz Materials Research Laboratory, University of Illinois at Urbana-Champaign, Urbana, USA

[5]Department of Physics, Hangzhou Normal University, Hangzhou, P. R. China

[6]Collaborative Innovation Center of Advanced Microstructures, Nanjing University, Nanjing, P. R. China

#these authors contribute equally to the work

*Email: yangliuphys@zju.edu.cn, hqyuan@zju.edu.cn

## Abstract

Combining angle-resolved photoemission spectroscopy and magneto-transport measurements, we systematically investigated the possible origin of the extreme magnetoresistance in Pr/Sm mono-antimonides/bismuthides (PrSb, SmSb, PrBi, SmBi). Our photoemission measurements reveal that the bulk band inversion and surface states are absent (present) in Pr/Sm antimonides (bismuthides), implying that topological surface states are unlikely to play an important role for the observed extreme magnetoresistance. We found that the electron-hole compensation is well satisfied in all these compounds and the bulk band structure exhibits no obvious temperature dependence from 10 K up to 150 K. Simultaneous fittings of the magnetoresistance and Hall coefficient reveal that the carrier mobility is dramatically enhanced at low temperature, which naturally explains the suppression of extreme magnetoresistance at high temperatures. Our results therefore show that the extreme magnetoresistance in these compounds can be well accounted for by the two-band model with good electron-hole compensation. Finally, we found that both PrSb and SmSb exhibit highly linear bulk bands near the $X$ point and lie close to the transition point between a topologically trivial and nontrivial phase, which might be relevant for the observed anomalous quantum oscillations.

# 1. INTRODUCTION

Recently rare-earth monopnictides, REPn (RE being a rare earth element, Pn being a group VI element), have attracted considerable research interest. These materials have a simple rocksalt structure and yet exhibit extreme magnetoresistance (XMR) at low temperature that does not show obvious saturation with the magnetic field [1,2]. More interestingly, the resistivity versus temperature under a large magnetic field exhibits a notable upturn and eventually plateaus at very low temperature, sharing great similarity with other topological materials such as $WTe_2$ [3,4], TaAs [5], NbP [6], and $SmB_6$ [7,8]. Since XMR is often found in topological materials with topological surface states (TSSs) [9,10,11,12,13], it is natural to link this peculiar transport property with possible TSSs in these systems. Within this picture, the XMR could be attributed to the suppression of backscattering of TSSs without a magnetic field, which is relaxed when a magnetic field is applied and breaks time-reversal symmetry [14,15,16]. Another possible explanation for the observed XMR is based on a classic model involving excellent electron-hole compensation and high mobility of charge carriers, resulting in XMR with a quadratic field dependence [17,18]. This point was supported by recent photoemission studies on some antimonides such as YSb, LaSb [19,20,21], where no TSS was found and yet good electron-hole balance was observed. However, in similar rare earth bismuthide LaBi, where XMR is also observed, large electron-hole imbalance was reported, casting doubts on this classic scenario [22]. Another possible contribution to XMR is the proposed *d-p* orbital mixing, although to what extent this mechanism could account for the XMR remains an open question [2].

In order to reveal the intrinsic connection between the TSSs and XMR, it is necessary to make systematic comparison of the magneto-transport properties and electronic structures for a set of closely related compounds with different electron configurations. This motivates us to study the Pr/Sm mono-antimonides/bismuthides, whose electronic structures could be varied due to differences in the *f* electron count and, more importantly, the strength of spin-orbit coupling (due to different Pn elements). Our study combines measurements from angle-resolved photoemission spectroscopy (ARPES), magneto-transport, and density functional theory (DFT) calculations, which could be important to develop a comprehensive understanding of the correlation between the magnitude of XMR, presence/absence of TSSs, and the extent of electron-hole compensation in the bulk band structure. Our work is also motivated by recent observations of anomalous quantum oscillations (QOs) in these materials, including field-dependent frequency change in PrSb

[23], obvious deviation of de Hass-van Alphen and Shubnikov-de Hass frequencies in SmSb [24]. These QO results imply that addition of *f* electrons from varied RE elements might modify the band structure, which could be further altered by the strong magnetic field. Nevertheless, direct measurements of the band structures and possible surface states are still lacking for PrSb/SmSb, hindering an in-depth understanding of these peculiar magneto-transport properties. More importantly, previous studies are focused on topological properties from either magneto-transport measurements on individual PrSb/SmSb compound [23,24], or energy-momentum dispersions of bulk and surface states in PrBi/SmBi [25], but there has been no systematic dataset and combinational analysis for the underlying origin of XMR, a key physical property for all these related compounds. The current paper represents a comprehensive effort towards this direction, by combining and analyzing all the relevant experimental results in a unified manner.

## 2. Experimental and Computational Details

Single crystals of PrSb, SmSb, PrBi and SmBi were grown using the flux method (the details of growth procedure could be found in [23]). ARPES measurements were carried out at BL13U beamline at National Synchrotron Radiation Lab in Hefei [26], and beamline 21B at Taiwan Light Source [27]. All ARPES spectra were taken with a R4000 electron energy analyzer at a sample temperature between 10 and 20 K (except Fig.5), i.e., the samples are all in their paramagnetic phase. The typical energy resolution is ~20 meV and the momentum resolution is typically 0.01 Å$^{-1}$. Large single crystals with a typical dimension of 1x1x1 mm were cleaved *in-situ* in the ARPES chamber at low temperature (base pressure < 9x10$^{-11}$ Torr), resulting in flat shiny surface for ARPES measurement. Sample aging is monitored regularly to ensure that the data presented in the paper reflects the intrinsic electronic structure of the materials.

Magneto-transport measurements were performed using a Quantum Design Physical Property Measurement System (PPMS). Four Pt wires were used for making the connection through Ag epoxy or spot welding, which was done immediately after cleaving the sample and exposing a fresh (001) surface. DFT calculations were performed using a plane-wave basis projected augmented wave method, as implemented in the Vienna Ab Initio Simulation Package (VASP) [28]. To ensure convergence, we employed plane-wave basis up to 480 eV, and 12x12x12 Γ-centered K-mesh so that the total energy converges to 1 meV per cell. Spin-orbit coupling is included in all calculations, and *f* electrons have been treated as core electrons. Since the Perdew

Burke and Ernzerhof (PBE) flavor of generalized gradient approximation to the exchange-correlation functional is known to exaggerate the band inversion features [29], we have employed a more accurate method, i.e., the modified Becke-Johnson (MBJ) potential [30]. To obtain the dispersion of TSSs, we used the surface Green's function method, assuming a semi-infinite slab. Specifically, the band structure from DFT calculations with MBJ potential were fitted to a tight-binding Hamiltonian using the maximally localized wannier function method with RE $t_{2g}$ orbitals and pnictogen-$p$ orbitals. The resulting Hamiltonian were then used to calculate the surface states using the surface Green's function.

## 3. Results and Discussion

### 3.1. Sample Characterization and XMR

Rare earth monopnictides share the simple rocksalt structure (face-centered cubic), with lattice constant of 6.38 Å, 6.46 Å, 6.27 Å and 6.36 Å for PrSb, PrBi, SmSb and SmBi, respectively. The crystal structure and the corresponding bulk three-dimensional (3D) Brillouin zone are shown in Fig. 1(a). Since only the in-plane momentum of electrons can be determined unambiguously by ARPES, the ARPES data are presented in the surface two-dimensional (2D) Brillouin zone, where the bulk $\Gamma$ and one $X$ ($k_z=\pi$) point project onto 2D $\overline{\Gamma}$ point with $(k_x,k_y)=(0,0)$, and two bulk symmetry-inequivalent $X$ points (with $k_z=0$ and $\pi$) project onto the same 2D $\overline{M}$ point. To confirm the sample quality of cleaved samples used for ARPES measurements, we performed momentum-integrated energy scans as shown in Fig. 1(b), which reveals sharp core level peaks expected for an impurity-free single crystal. The sample quality is also confirmed by the high residual resistivity ratio, typically >100 for all four compounds studied in this paper. XMR was observed for all these compounds – Fig. 1(c) shows the results for PrSb and PrBi, where the resistivity is always metallic in the absence of magnetic field and becomes insulating under a large magnetic field, eventually plateauing at very low temperature. We define the magnetoresistance (MR) as

$$MR = \frac{\rho(B) - \rho(B=0)}{\rho(B=0)}, \quad (1)$$

and summarize the MR values from our best samples at 2K up to 9T for all four compounds in table I. The resistivity vs magnetic field at low temperature is summarized in Fig. 1(d); the resistivity clearly exhibits a quadratic field dependence and shows no sign of saturation up to 9 T (see inset in Fig. 1(d)). To further illustrate the magnetic field dependence, we plot

$log(MR)$ vs $log(B)$ at different temperatures for SmSb in Fig. 1(e). The similar slope found in different temperatures implies that MR likely shares similar origin in the wide temperature range, as we shall discuss in detail below.

*3.2 Band Topology and TSSs*

Figure 1(f) shows the calculated bulk band structure along the $\Gamma X$ direction for PrSb, using the DFT calculations with the MBJ potential. The bands near the Fermi level mainly come from Pr 5*d* and Sb 5*p* orbitals, as indicated by the different colors in Fig. 1(f). Near the *X* point, the Pr 5*d* band crosses the Sb 5*p* band, creating an inverted (partial) band gap and resulting in nontrivial band topology. According to the topological theory, the $Z_2$ topological invariant can be defined and calculated in REPn [31,32,33]: the band inversion between RE 5*d* orbital and Pn *p* orbital near the *X* point would give rise to nontrivial $Z_2$ topological invariant and TSSs; if no band inversion occurs near *X*, the system would be topologically trivial without TSSs.

Before discussing the ARPES results, we first take a look at the results obtained from DFT MBJ calculations [33], summarized on the right panels in Fig. 2. The projected bulk bands are shown as continuous blue background and the bulk band edges can be clearly identified. Near the $\overline{M}$ point, the calculations predict a clear inverted band gap for PrSb, PrBi and SmBi, within which the bulk spectral weight is absent. Meanwhile, sharp peaks exist within this inverted partial gap, which should correspond to the theoretically predicted TSSs [34,35]. For SmSb, the bulk Sm 5*d* and Sb 5*p* bands almost touch near the $\overline{M}$ point (see the white arrow), indicating that it could be close to a transition point between topologically trivial and nontrivial phase [33]. On the left panels of Fig. 2, we show the corresponding experimental ARPES spectra taken with 28.1 eV photons. While the PrBi and SmBi data is adapted from Ref. [25], the data from PrSb and SmSb has not been reported before. Due to a large $k_z$ broadening for this class of materials [36,37,38], the ARPES spectra near this photon energy are found to contain contributions from a broad range of $k_z$ cuts, including both $k_z=0$ and $k_z=\pi$. As a result, the experimental results can be directly compared with DFT calculations using the surface Green's function method (right panels in Fig. 2), which essentially consist of equal spectral contribution from all $k_z$ cuts. For PrBi and SmBi, the agreement between experiments and DFT calculations is excellent, showing clear bulk band inversion and surface states (SSs) lying inside the gap. We mention here that the two originally gapless TSSs near the $\overline{M}$ point interact to yield two sets of gapped SSs, which shows up in both PrBi and SmBi data (see [25] for detailed discussions/analysis). On the other hand, for PrSb, the predicted bulk band inversion is absent in the experimental data, which shows no obvious sign of inverted bulk gap and SSs. This is consistent with the zero Berry's phase obtained in QO measurements [23]. For SmSb, the experimental results show that the Sm 5$d$ and Sb 5$p$ bands become very close near the $\overline{M}$ point, consistent with the DFT calculations, although no obvious sign of band inversion can be observed. Note that DFT calculations predict a large change of band structure from PrSb to SmSb [33], while the actual change from experiments is much less. This discrepancy is likely due to inaccuracy in DFT calculations for these low carrier density systems [18]. Another possible explanation is the correlation effect due to $f$ electrons in the system, which might be relevant at low temperatures and is ignored in the current DFT calculations assuming completely local $f$ electrons. We also ignore any possible small valence change for Pr/Sm at the surface: the cleaved (001) surface is charge neutral and nonpolar, and therefore no significant charge redistribution would be expected.

Since all four compounds exhibit XMR of similar magnitude (see table I), the absence (presence) of SSs in antimonides (bismuthides) indicate that SSs do not play a major role in XMR. This is not surprising as the observed SSs are not close to the Fermi level and therefore probably do not contribute considerably to the transport properties. Recent studies on LaSb and YSb suggest that the classic model of good electron-hole compensation is more likely to account for the XMR in these systems [19, 20]. We now test this idea and make quantitative comparison on these four compounds.

*3.3 Electron-hole Compensation*

Although ARPES reveals three bulk bands contributing to the Fermi surface for these compounds, we use an effective two-band model to analyze the MR and Hall coefficient. This simplification is based on the following consideration: the two hole bands are derived from the same Pn $p_{3/2}$ orbitals, and therefore their mobility is probably close (hence their contributions to magneto-transport could be factorized into one single hole band). The simplified treatment in the two-band model also help test the electron-hole compensation in a more reliable manner, as we shall discuss later. Within this model, the resistivity $\rho$ can be described as [17]

$$\rho(B) = \frac{(n_h\mu_h + n_e\mu_e) + (n_h\mu_e + n_e\mu_h)\mu_h\mu_e B^2}{e(n_h\mu_h + n_e\mu_e)^2 + e(\mu_h\mu_e)^2(n_h - n_e)^2 B^2}, \tag{2}$$

where $B$ is the magnetic field, $n_e$ ($n_h$) is the carrier density of electrons (holes), and $\mu_e$ ($\mu_h$) is the mobility of electrons (holes), respectively. As a result, the magnetoresistance, MR, can be rewritten as

$$\text{MR} = \frac{\rho(B) - \rho(0)}{\rho(0)} = \frac{n_h \mu_h n_e \mu_e (\mu_h + \mu_e)^2 B^2}{(n_h \mu_h + n_e \mu_e)^2 + (n_h - n_e)^2 (\mu_h \mu_e)^2 B^2}. \tag{3}$$

Based on eq. (3), there are two prerequisites for XMR: Excellent electron-hole compensation ($n_e \sim n_h$) and very high mobility of charge carriers.

Theoretically, the carrier density can be estimated by integrating the enclosed volume of occupied bulk bands at the Fermi level in the 3D momentum space. The calculated 3D Fermi surfaces of these materials, shown in Fig. 3(a,b) for PrSb and PrBi, consist of two hole bands at the $\Gamma$ point (corresponding to β, δ orbitals observed in QO measurements, see experimental 2D Fermi surface in Fig. 3(c,d)) and one elliptical electron band at each $X$ point (there are three inequivalent $X$ points). Projections from electron bands at different $X$ points result in a small circular band at $\overline{\Gamma}$ point (α orbital) and two elliptical bands at $\overline{M}$ point (marked as γ and γ') that are rotated 90° with respect to one another. All these orbitals can be observed in the experimental 2D Fermi surfaces in Fig. 3(c,d). In particular, the high intensity pocket γ, indicated by circles, is from $k_z=0$ cut, while the low intensity pocket γ' is from $k_z=\pi$ cut. Due to the aforementioned large $k_z$ broadening [25], it becomes straightforward to extract the bulk band edges by simply tracing out the outer boundaries of continuous bulk bands, shown as open circles in Fig. 3(c,d). The corresponding area for each band is summarized in table II for all four compounds. Although these values are in reasonable agreement with those obtained from QOs [23,24], some quantitative differences are present, whose origin is not understood at the moment. More discussions on this will be presented in Sec. 3.5.

Based on the extracted bulk band edges, one could estimate the 3D carrier density by applying the cubic lattice symmetry and integrating the enclosed volume of occupied bands in 3D momentum space. While our detailed analysis shows that the carrier density ratio ($n_e/n_h$) is reasonably close to 1 for all four compounds, a large error bar from such analysis might be present. This is because the bulk band edges are usually broad and difficult to determine with high precision. A more reliable method for extracting the carrier density ratio is based on simultaneous fittings of MR (eq. (3)) and Hall coefficient. Here the Hall coefficient can be written as

$$\rho_{xy}(B) = \frac{B}{e} \frac{(n_h\mu_h^2 - n_e\mu_e^2) + (n_h - n_e)(\mu_h\mu_e)^2 B^2}{(n_h\mu_h + n_e\mu_e)^2 + (n_h - n_e)^2(\mu_h\mu_e)^2 B^2}. \qquad (4)$$

Since such fitting requires only four fitting parameters (for two separate curves), the results can be quite reliable. An example of such analysis is shown in Fig. 4(a,b) for SmSb, where good fittings for both MR and Hall coefficients can be obtained in a wide temperature range, indicating that the two-band model captures the essential physics in these systems. The extracted carrier density and mobility is summarized in table I for PrSb, PrBi, SmSb and SmBi at 2 K. Our results clearly show that the electron-hole compensation is well satisfied for all four compounds at low temperature. This result is consistent with the intuitive picture that the covalent bonding between RE 5*d* and Pn *p* orbitals results in an equal number of charge carriers from both bands at the Fermi level, i.e., a compensated semi-metal. This is also in line with the quadratic field dependence of the resistivity (Fig. 1(d)), which can be well understood from Eq. (2) assuming $n_e \sim n_h$. The fittings also yield a large carrier mobility at low temperature, in the order of $10^4$ cm$^2$/V·s, which is essential for the observed XMR.

It should be pointed out that there is a slight deviation from the quadratic field dependence in the MR at the low magnetic field, as shown in Fig. 1(d). This is because the magnetic scattering from the local moments, which is not included in eq. (2,3), could change significantly at low magnetic field when the local moments are polarized according to the external field. This effect is most obvious at lower field, when the local moments just start to polarize and the MR from electron-hole compensation (based on eq. (2)) does not dominate.

*3.4 Temperature Dependence*

An outstanding question of XMR observed in these materials is why XMR is only present at very low temperature. If the electron-hole compensation scenario is true, the suppression of XMR at high temperature could be caused by either imbalance of electron-hole compensation or significantly reduced charge carrier mobility, as indicated in eq. (3). To investigate whether the electron-hole imbalance occurs at high temperature, we performed temperature-dependent ARPES measurements – an example for SmSb is shown in Fig. 5. We observe no obvious change in the pocket size of the bulk bands over a wide temperature range (from 10K up to 150 K), where significant change in MR already takes place (Fig. 4(a)). Similar temperature dependence is observed in other compounds, indicating that the electron-hole balance is not dependent much on the temperature.

Therefore, if electron-hole compensation is the underlying mechanism for XMR, its suppression at high temperature is most likely attributed to the significant reduction of the carrier mobility. In the standard transport theory, the carrier mobility $\mu$ is defined by

$$\mu = \frac{e}{m^*}\tau, \tag{5}$$

where $e$, $m^*$ is the carrier charge and effective mass, $\tau$ is the average scattering time, which is inversely proportional to the scattering rate. In an ordered 2D system, where the photoemission linewidth is mainly determined by the lifetime broadening of the initial state, the scattering rate of charge carriers can be estimated by the width of quasiparticle peaks in the momentum space. However, in a 3D system such as REPn, the photoemission linewidth is affected by the lifetime broadenings of both the initial and final states [39], i.e., the photoexcitated final state could also make a significant contribution to the peak linewidth. This is evident from the large linewidth of the bulk bands in the experimental momentum distribution curves shown in Fig. 5(c-d), which is $\sim 5\times10^{-2}$ Å$^{-1}$, much larger than the estimated broadening of $\sim 1\times10^{-4}$ Å$^{-1}$ from the resistivity data (see Appendix 1 for a detailed estimation). This indicates that the photoemission linewidth is dominated by the final state effect, and it seems unlikely to resolve such a small reduction in the scattering rate from ARPES peak width due to overwhelming contributions from the final state effect. This naturally explains the absence of any change in the linewidth with the temperature in Fig. 5(d).

Given the very low resistivity (and high mobility) in these systems, the magneto-transport measurement is a much more sensitive gauge of the scattering rate of the charge carriers than ARPES. As such, we perform similar fittings of MR and Hall coefficients for SmSb in a wide temperature range, and plot the carrier mobility and density as a function of temperature in Fig. 4(c,d). Our fittings reproduce the experimental results very well (Fig. 4(a,b)), including the quadratic field dependence of MR in a large temperature range (also see Fig. 1(e)) and the non-linear field dependence of Hall coefficient. The deduced carrier mobility indeed exhibits dramatic decrease with increasing temperature, which could lead to the disappearance of XMR at high temperature. Our fittings also show that the electron-hole compensation is reasonably satisfied in a wide temperature range, consistent with our ARPES measurements. We note that a small imbalance of electron-hole carrier density ($n_e/n_h$~1.1) above 50 K is suggested based on Fig. 4(c,d), which could indicate that a slight mismatch of carrier density might make additional contribution to the suppression of XMR at high temperature. However, such a carrier density change, which translates to ~3% change in the Fermi vector, is not observed in ARPES data. One possibility is that such a small vector change might not be resolvable in ARPES spectra due to the large width of bulk bands in momentum space. Another more likely explanation is that the two-band simplification in these three-band systems could generate somewhat larger uncertainty/error in high temperature fittings due to the diminishing MR, resulting in an overestimate of the electron-hole imbalance.

*3.5 Effective Mass and Linear Bulk Bands*

Based on discussions above, the XMR in these compounds is achieved through high carrier mobility with compensated hole/electron carriers. According to eq. (5), a small effective mass is preferable to achieve high mobility (and hence XMR). The effective mass can be estimated based on the following classic equation

$$m^* = \frac{\hbar k_F}{v_F}, \tag{6}$$

where $k_F$ and $v_F$ is the crystal momentum and group velocity of the charge carrier at the Fermi level. Both $k_F$ and $v_F$ can be directly obtained from the ARPES spectra, which allows us to calculate the effective mass for each individual band, as summarized in table II. These values are comparable to the results obtained from QO measurements [23], attesting the validity of this analysis. In particular, the small hole pocket (β) and electron pocket (α) possess very small effective mass (~0.2 $m_e$), which likely make important contributions to the XMR in these systems. Note that the effective mass is highly anisotropic for the elliptical electron pocket at *X* point, with almost an order-of-magnitude difference along two different directions (compare α and γ orbitals in table II). How such an anisotropic band in a cubic lattice affects the magneto-transport properties will be an interesting question for future studies.

Another peculiar feature in the band structures of PrSb and SmSb is their remarkably linear electron band at *X* point (Fig. 6), which is essentially the α orbital and plays a dominant role in the QO measurements [23]. Their linear dispersion can be best seen in the energy-momentum dispersion cuts (a,b), resulting in a circular shape in the constant energy contours over a large energy range (see SmSb data in Fig. 6(c)). Similar dispersion has also been reported for other RESb compounds [40,41]. Such an anomalous linearity in this electron band is likely due to proximity to a transition point between a topologically trivial and nontrivial phase, as discussed in Sec. 3.2. Specifically, ARPES spectra in Fig. 2 show that the RE $5d$ and Sb $5p$ bands become very close near the bulk $X$ point, although they do not actually cross to form the band inversion required for TSSs. As a result, the RE $5d$ and Sb $5p$ bands develop highly linear dispersion near the $X$ point, close to those observed in a Dirac semimetal (although a small gap persists due to lack of symmetry protection) [42,43]. The proximity between RE $5d$ and Sb $5p$ bands also indicates that a weak perturbation to the band structure, e.g., a magnetic field, could possibly induce a band inversion in some cases and render the system topologically nontrivial. Experimentally, a field-induced change in QO frequencies is indeed observed in PrSb [23], although the residual Landau index remains close to zero up to 32 T. On the other hand, for SmSb, a half-integer residual Landau index is obtained under a large magnetic field [24], implying possible field-induced nontrivial band topology. Although the different field dependence between PrSb and SmSb is not understood at the moment, it might be attributed to different $f$ electron configurations, resulting in different couplings with the valence/conduction electrons at low temperature (and large magnetic field).

## 4. Conclusions

To conclude, we have performed systematic ARPES and magneto-transport measurements/analysis on Pr/Sm mono-antimonides and bismuthides, which provide important insight to understand the origin of XMR observed in these materials. We found that the bulk band inversion and SSs are absent (present) in the antimonides (bismuthides). DFT calculations using the MBJ potential show good agreement with ARPES results for the bismuthides, but significant deviations occur for the antimonides. The strong XMR observed in antimonides, despite the absence of any TSS, clearly show that TSSs do not play a major role in the observed XMR. Instead, our simultaneous fittings of the MR and Hall coefficient reveal that the electron-hole compensation is well satisfied in a wide temperature range, while the largely enhanced carrier mobility at low temperature could be key for the observed XMR. This is consistent with our temperature dependent ARPES measurements, although the reduced carrier scattering at low temperature cannot be resolved in ARPES spectra due to dominant momentum broadening from final state effects. In general, our combined analysis of ARPES and magneto-transport results demonstrate that the observed XMR can be very well explained by electron-hole compensation in the two-band model, which gives rise to the temperature-independent power law $MR \propto B^2$ (Fig. 1(e)). For PrSb and SmSb, our results also show that both compounds are close to the transition point from a topologically trivial to nontrivial phase, which results in peculiar bulk bands with highly linear dispersion. Such insight is made possible through a combinational analysis of both magneto-transport and ARPES results, which has not been performed in earlier studies.

**Acknowledgments**

This work is supported by National Key R&D Program of the MOST of China (Grant No. 2017YFA0303100, 2016YFA0300203), National Science Foundation of China (No. 11674280, No. 11274006), and the Science Challenge Project of China (No. TZ20160004). TCC acknowledges support from the US Department of Energy under Grant No. DE-FG02-07ER46383. We would like to thank Pengdong Wang, Yi Wu, Dr. Chanyuen Chang, Prof. Deng-sung Lin for help during synchrotron ARPES measurements, and Prof. Haijun Zhang, Prof. Yi Zhou, Prof. Frank Steglich for helpful discussions.

**Appendix 1**

In this appendix, we show how the momentum broadening from the scattering of the initial state can be estimated from the electrical resistivity. Taking SmSb as an example, the typical resistivity is ~40 μΩ cm at 100 K. Now use

$$\rho = \frac{m^*}{ne^2\tau}, \tag{A1}$$

where $m^*$ is the effective mass of carrier, $n$ is the carrier concentration, $e$ is electron charge and $\tau$ is the carrier lifetime. For simplicity, we let $m^* \sim 0.75 m_e$ (the averaged value of charge carrier mass based on table II). $n$ is approximately $2 \times 10^{20}$ cm$^{-3}$, based on transport measurements. Now based on eq. (A1), we can estimate the carrier lifetime $\tau$ to be $\sim 10^{-12}$ s. Next, we use the uncertainty principle

$$\Delta E \sim \frac{\hbar}{\tau} \sim 10^{-3} \text{ eV}, \tag{A2}$$

together with a group velocity of ~4 eV Å for the bands near $E_F$ (from ARPES data), we can estimate the initial state contribution to the MDC width on the order of $\Delta k \sim 10^{-4}$ Å$^{-1}$.

|      | MR (%) at 9 T | $\mu_e (cm^2/Vs)$ | $\mu_h (cm^2/Vs)$ | $n_e (cm^{-3})$ | $n_h (cm^{-3})$ | $n_e/n_h$ |
|------|---------------|-------------------|-------------------|-----------------|-----------------|-----------|
| PrSb | $1.32 \times 10^4$ | $2.28 \times 10^4$ | $2.80 \times 10^4$ | $1.26 \times 10^{20}$ | $1.27 \times 10^{20}$ | 0.99 |
| PrBi | $0.95 \times 10^4$ | $0.93 \times 10^4$ | $0.77 \times 10^4$ | $1.83 \times 10^{20}$ | $1.81 \times 10^{20}$ | 1.01 |
| SmSb | $1.38 \times 10^4$ | $0.51 \times 10^4$ | $0.52 \times 10^4$ | $2.42 \times 10^{20}$ | $2.44 \times 10^{20}$ | 0.99 |
| SmBi | $2.90 \times 10^4$ | $2.06 \times 10^4$ | $1.72 \times 10^4$ | $3.60 \times 10^{20}$ | $3.60 \times 10^{20}$ | 1.00 |

Table I. A summary of MR, carrier mobility ($\mu_e$, $\mu_h$), carrier density ($n_e$, $n_h$) and electron-hole compensation in PrSb, PrBi, SmSb and SmBi (from our best samples). All the data are obtained from magneto-transport measurements at 2 K. The carrier mobility and density are derived from simultaneous fittings of the MR and Hall coefficients based on the two-band model (see Fig. 4).

|      | Fermi surface area($Å^{-2}$) | | | | $m^*(m_e)$ | | |
|------|--------------|--------------|--------------|--------------|--------|--------|--------------------|
|      | Small hole($\beta$) | Large hole($\delta$) | Electron($\alpha$) | Electron($\gamma$) | Hole $\beta$ | Hole $\delta$ | Electron($\alpha$, $\gamma$) |
| PrSb | 0.0331 | 0.0648 | 0.0201 | 0.0648 | 0.19 | 0.59 | 0.13, 2.32 |
| PrBi | 0.0445 | 0.1384 | 0.0142 | 0.0927 | 0.17 | 1.20 | 0.18, 1.80 |
| SmSb | 0.0383 | 0.1280 | 0.0254 | 0.0910 | 0.21 | 0.73 | 0.17, 2.69 |
| SmBi | 0.0627 | 0.1828 | 0.0201 | 0.1222 | 0.24 | 1.20 | 0.17, 1.30 |

Table II. A summary of Fermi surface area and effective mass for all bands, obtained from ARPES measurements. Here we use β, δ, α, γ labels to facilitate direct comparison with QO measurements (see Fig. 3 and [23,24]). Note that α and γ orbitals are essentially different cross cuts of the same bulk electron band at *X* point.

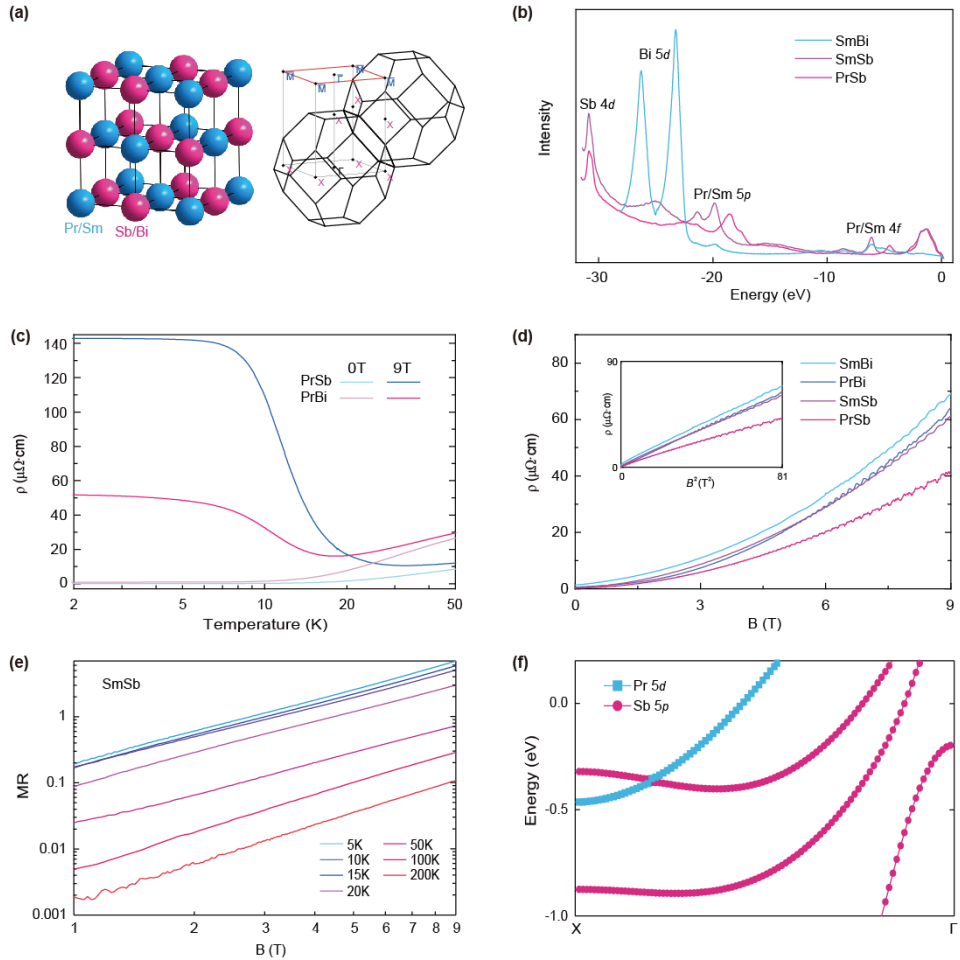

Fig. 1 (Color Online). Sample characterization and XMR. (a) Crystal structure (left) and 3D bulk Brillouin zones with projected 2D surface Brillouin zone typically used in ARPES measurements (right). (b) Momentum-integrated large-range energy scans, showing the expected sharp core level peaks. The spectra for SmBi were taken at 54 eV, while all other spectra were taken at 42 eV. (c) The resistivity (ρ) vs temperature with and without the magnetic field for PrSb and PrBi. (d) Resistivity vs magnetic field (B) at 2 K for all compounds studied in the paper. The ripples at high field arise from QOs in these materials. Inset is a plot of the same data but plotted with $B^2$, highlighting a quadratic field dependence of the MR. (e) MR as a function of the magnetic field for SmSb, plotted in the log scale at various temperatures. (f) Calculated bulk band structure along the $\Gamma X$ direction for PrSb, showing the predicted bulk band inversion between Pr 5$d$ (green) and Sb 5$p$ (red) bands near the $X$ point.

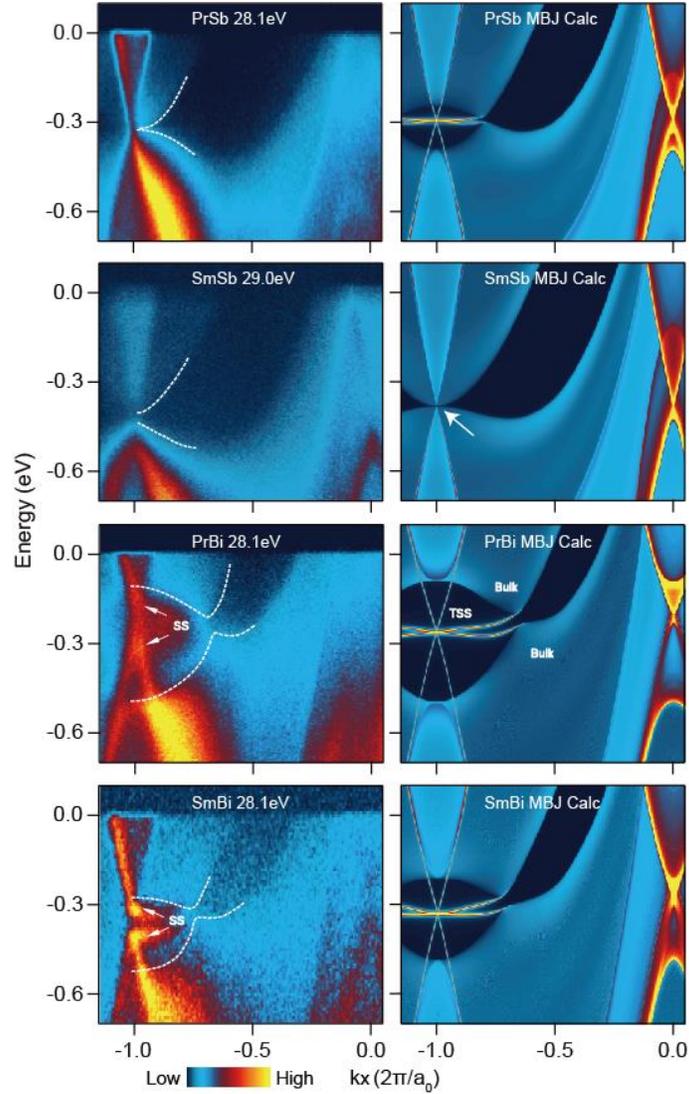

Fig. 2 (Color Online). Band structures of PrSb, SmSb, PrBi and SmBi along the $\overline{M}\,\overline{\Gamma}$ direction in the 2D surface Brillouin zone ($a_0$ is the respective lattice constant for the conventional unit cell). Left are experimental ARPES spectra, and right are results from DFT calculations with the MBJ potential, using the surface Green's function method. The white dashed lines on top of the experimental data indicate the extracted bulk band edges, while the white arrows denote the SSs. In the calculation, the blue background indicates the bulk continuum and the TSSs are shown as sharp states lying within the inverted gap (see labels in the PrBi calculation). The white arrow in the calculation of SmSb highlights the crossing point between Sm $5d$ and Sb $5p$ bands.

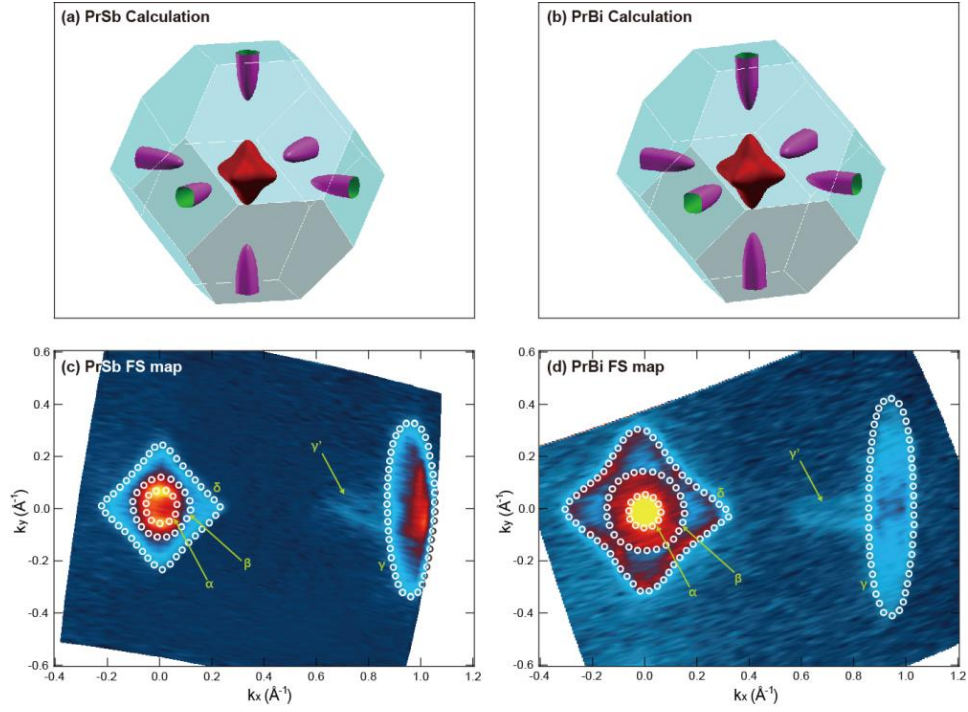

Fig. 3 (Color Online). (a,b) Calculated 3D Fermi surface for PrSb (a) and PrBi (b). (c,d) Experimental 2D Fermi surface of PrSb (c) and PrBi (d). The circles indicate the extracted momentum positions of bulk band edges, where were used to calculate the Fermi surface area in table II. Due to a large $k_z$ broadening, contributions from both $k_z=0$ (β, δ, γ' pockets) and $k_z=\pi$ cuts (α, γ pockets) are present in the maps. Note that γ and γ' pockets are rotated by 90° with respect to one another, and they originate from different $k_z$ cuts of the symmetry-equivalent electron bands.

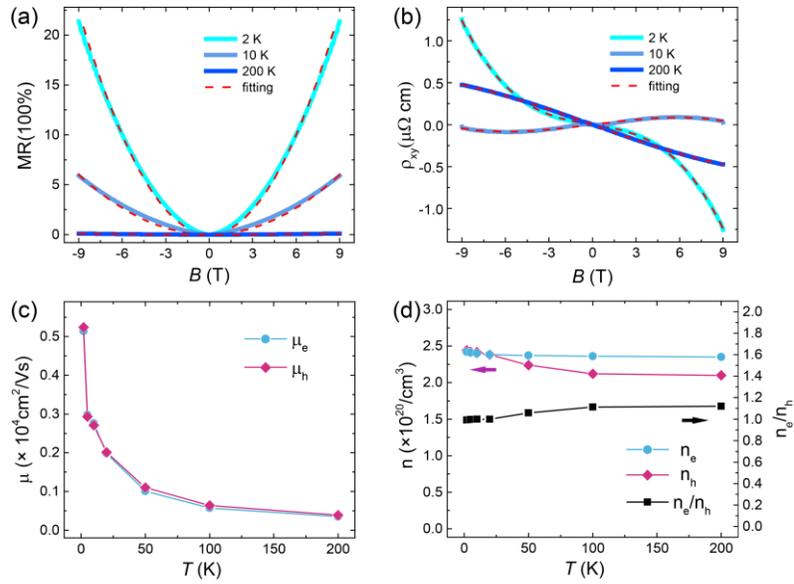

Fig. 4 (Color Online). Extracting the carrier density and mobility for SmSb by simultaneously fitting the MR and Hall data. (a,b) Field dependence of MR (a) and $\rho_{xy}$ (b) at 2K, 10K and 200K with fits (red dash line). (c,d) Extracted carrier mobility (c) and density (d) in a large temperature range.

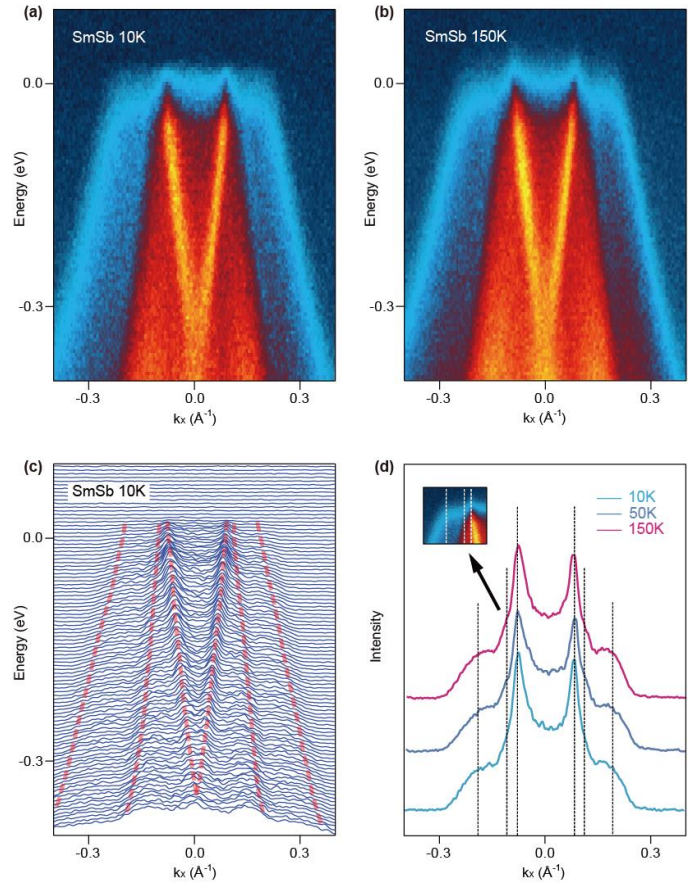

Fig. 5 (Color Online). Temperature evolution of the SmSb band structure. (a-b) ARPES Spectra of SmSb near the $\bar{\varGamma}$ point taken with 23.8 eV photon at 10 K (a) and 150 K (b), respectively. (c) Waterfall plot of Momentum distribution curves at 10 K. Red dashed lines are guide to eyes for the dispersions of two hole bands and one electron band near the Fermi level. (d) Momentum distribution curves at the Fermi level at a few representative temperatures. Vertical dashed lines indicate the momentum positions for two hole bands and one electron band. Inset is an enlarged view of spectra near Fermi level. No obvious temperature dependence in the peak position and width can be observed.

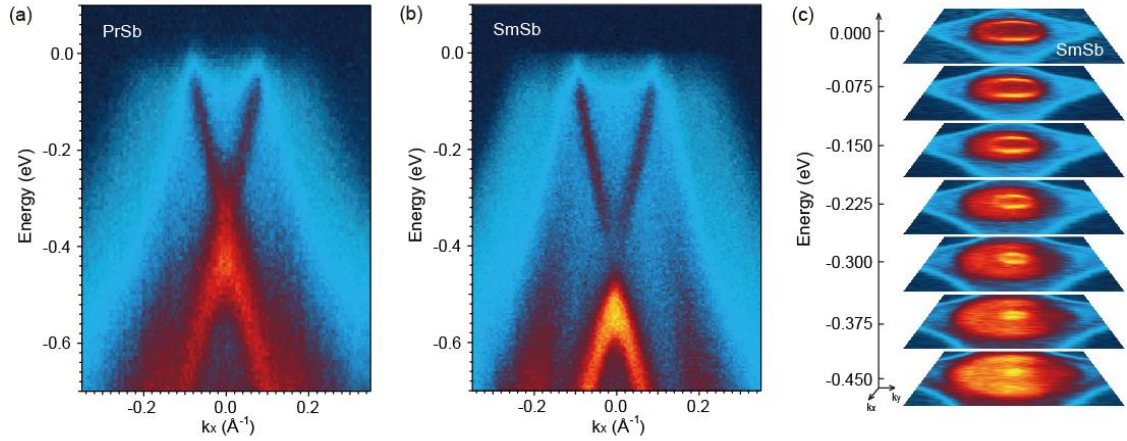

Fig. 6 (Color Online). (a,b) Energy-momentum cuts near the $\bar{\Gamma}$ point for PrSb and SmSb, respectively, showing the highly linear electron bands (α orbital). (c) Constant energy contours for SmSb at various energies near the $\bar{\Gamma}$ point. Due to the large $k_z$ broadening, the ARPES spectra consist of contributions from two hole bands from bulk $\Gamma$ point ($k_z=0$, corresponding to δ, γ orbitals) and one electron band at bulk $X$ point ($k_z=\pi$).